\documentclass[aps,prl,preprint,groupedaddress]{revtex4-1}

\usepackage{graphicx}
\usepackage{rotating}
\usepackage{amssymb}    
\usepackage{amsmath}   
\usepackage{epsfig}
\usepackage[normalem]{ulem}   
\usepackage{bm}   
\usepackage{color}

\date{\today}

\begin{document}

\title{Trapping Phenomenon Attenuates Tipping Points for Limit Cycles}

\author{Everton S. Medeiros}
\email{esm@if.usp.br}
\affiliation{Institute of Physics, University of S\~ao Paulo, S\~ao Paulo, Brazil}
\author{Iber\^e L. Caldas}
\affiliation{Institute of Physics, University of S\~ao Paulo, S\~ao Paulo, Brazil}
\author{Murilo S. Baptista}
\affiliation{Institute for Complex Systems and Mathematical Biology, SUPA, University of Aberdeen, Aberdeen, United Kingdom}
\author{Ulrike Feudel}
\affiliation{Institute for Chemistry and Biology of the Marine Environment, Carl von Ossietzky University Oldenburg, Oldenburg, Germany}

\begin{abstract}

Nonlinear dynamical systems may be exposed to tipping points, critical thresholds at which small changes in the external inputs or in the system’s parameters abruptly shift the system to an alternative state with a contrasting dynamical behavior. While tipping in a fold bifurcation of an equilibrium is well understood, much less is known about tipping of oscillations (limit cycles) though this dynamics are the typical response of many natural systems to a periodic external forcing, like e.g. seasonal forcing in ecology and climate sciences. We provide a detailed analysis of tipping phenomena in periodically forced systems and show that, when limit cycles are considered, a transient structure, so-called channel, plays a fundamental role in the transition. Specifically, we demonstrate that trajectories crossing such channel conserve, for a characteristic time, the twisting behavior of the stable limit cycle destroyed in the fold bifurcation of cycles. As a consequence, this channel acts like a ``ghost” of the limit cycle destroyed in the critical transition and instead of the expected abrupt transition we find a smooth one. This smoothness is also the reason that it is difficult to precisely determine the transition point employing the usual indicators of tipping points, like critical slowing down and flickering.

\end{abstract}

\maketitle

\section{Introduction}

Many systems in nature possess a multitude of coexisting stable states for a given set of parameters reflecting environmental conditions. This phenomenon called multistability has been studied for decades because of its dynamical complexity arising from the coexistence of the different states (for a review cf. \cite{Feudel2008} and references therein). Examples can be found in various disciplines of science, such as pattern recognition in neuroscience \cite{Hertz1991,Canavier1993,Foss1996}, nonlinear optics with different phenomena in lasers \cite{Arecchi1982,Wieczorek2002} and coupled lasers \cite{Prengel1994,Pisarchik2002} or manifested in different outcomes in chemical reactions despite large care taken to realize the same initial conditions \cite{Hudson1981,Ganapathisubramanian1984}. Since the multitude of coexisting states can usually be related to different performances of the system, various control strategies have been developed to gear the system from one state to another in a prescribed way or to avoid states which correspond to undesired behavior of a system (for a review cf. \cite{Pisarchik2014} and references therein).  
%So-14

More recently, during the last decade, the study of systems possessing only two alternative stable states has gained increasing interest due to their importance particularly in climate science and ecology (cf. reviews \cite{Lenton2011,Folke2004,Scheffer2015} and references therein). It has been recognized that one of the great challenges to science consists in understanding critical transitions or shifts in dynamics or properties arising in natural systems as a response to global change. Such transitions, in ecology often called catastrophic or regime shifts, are in general related to either changes in the dominance of particular species resulting in different ecosystem services or even in loss of biodiversity \cite{Scheffer2001,Nes2007}. More specific alternative stable states have been identified in several limnic and marine ecosystems such as kelp forests \cite{Steneck2002}, coral reefs  \cite{Hughes1994, Knowlton2004387}, shallow lakes \cite{Scheffer2001}, seagrass meadows \cite{Carr2010}, where the alternatives are usually between the dominance of the native species like kelp, corals or seagrass and the undesired states dominated by algae. Also in terrestrial ecosystems such as the Sahara \cite{Koppel1997} or more general semiarid ecosystems \cite{Kefi2007}, in which the two alternative stable states are a vegetated and a desert state. 

In climate science, where these transitions are often called tipping points, several components of the climate system have been identified to be possibly vulnerable with respect to certain perturbations \cite{Lenton2011,Lenton2012}. Such tipping elements are related to several climate phenomena such as ElNino-Southern oscillation \cite{Lenton2011}, the Indian Monsoon \cite{Zickfeld2005}, the arctic and antarctic ice covers \cite{Lenton2011}. Additionally, as one of the first climate components, the thermohaline ocean circulation has been found to possess two alternative stable states, one of which related to the present climate with a large transport of heat to the Northern latitudes and the other one corresponding to a shut-down of the circulation consequently ceasing the heat transport \cite{Rahmstorf2001}. As a part of the thermohaline ocean circulation, local deep ocean convection is also vulnerable to a shut down \cite{Lenderink1994,Rahmstorf2002}. Both processes, the shut-down of THC on a global scale and deep-ocean convection on a local scale would have a large impact on the climate in the Northern hemisphere leading finally to a cooling in Northern- and Western Europe. The Indian Monsoon is expected to become more wet or more dry depending on which of the processes responsible for such changes like increasing albedo due to aerosol concentration or stronger El Nino's, respectively, are dominant in the future \cite{Zickfeld2005}. Another highly debated tipping point relates to the tendency of the arctic ice sheets to become thinner and finally to lead to an ice-free state in summer \cite{Eisenman2009, Notz2009}. Finally, we mention the different approaches to study the recurrent switchings of ice ages and warmer climates before the Holocene, which are attributed to several stable states and transitions between them \cite{Paillard2006}.

Different scenarios are discussed in literature that lead to such critical transitions. On the one hand, changes in the environmental forcing, e.g. atmospheric temperatures or altered precipitation patterns can induce such transitions by reaching a critical threshold at which one of the states loses its stability and the system switches to another state. In mathematical terms, this scenario is related to a bifurcation; the combination of two such bifurcations often comprises a hysteresis \cite{Guckenheimer2013} allowing for a switching between two alternative states, when a control parameter is varied \cite{Scheffer2001}. When two stable states coexist, then a switching between them is mediated by fluctuations leading to noise-induced transitions \cite{Horsthemke1984,Scheffer2009}.   

Due to the increasing concerns about such critical transitions on our planet earth, there is an urgent need to identify the approach of a regime shift or a tipping point before its occurrence. Such early warning signals, if applicable, can be used to anticipate the transition and to take measures to slow down the approach in the worst case or to avoid it if the expected alternative state is for some reason undesired. Such avoidance might not always be possible, particularly not in the climate system, but early warnings could be used to take political actions. During the last decade several methods have been developed to gain more insights into how to predict abrupt changes in the system dynamics, induced by its nonlinearity.One of the earliest measures identified is related to the time which is needed to respond to perturbations. While far away from the critical threshold, such perturbations die out quite quickly, this damping becomes significantly slower in the neighborhood of a threshold \cite{Horsthemke1984,Scheffer2009}. The perturbation applied can be either a single perturbation or some noise which is inevitable in experiments and in natural processes. In case of a single prescribed perturbation, this measure is easy to implement in experiments and therefore widely used in quantifying the distance to the threshold value.
%\cite{Hillebrand}. 

In case of a noisy system the approach of the critical threshold can be quantified by an increasing variance and autocorrelation \cite{Scheffer2009}. As a second effect noise leads to an irregular switching process between the two (or more) alternative stable states. This process is called flickering \cite{Scheffer2009,Wang2012}, attractor hopping \cite{Kraut1999} or chaotic itinerancy \cite{Kaneko2003, Masoller2002} depending on the context in which it is studied. Hence, a second indicator has been introduced measuring this flickering or hopping process, which occurs in a bistable (or multistable) region in parameter space in which two or more stable states coexist. It is important to note that the hopping dynamics depends on the noise strength. While a large body of work is devoted to the impact of additive noise, many processes in nature, particularly in ecosystems, are affected by multiplicative noise, which has only rarely been considered. It has to be emphasized that environmental noise in ecosystem dynamics is always multiplicative and plays by far the more important role \cite{Hastings2010,Sharma2015}. However, most of the previous work on indicators for critical thresholds is restricted to additive noise describing the impact of fluctuations on physical processes in the climate system, but being of limited value for ecological problems.  

Many bistable systems considered in nature possess two stable equilibria, i.e. the system is stationary. For the above mentioned example of shallow lakes, the water in the lake loses transparency by shifting abruptly from a clear to a turbid state when a threshold in the level of nutrients is reached. As a result of this eutrophication process, submerged plants dramatically disappear beyond a tipping point \cite{Scheffer2001}. For the example of desertification, rainfall patterns are the essential environmental conditions determining the shift from a perennial vegetation via localized vegetation patterns to the state of bare soil. Moreover, taking the grazing pressure by livestock in the Sahel zone into account, leads again to a shift from a perennial to an annual vegetation \cite{Koppel1997}. However, in analyzing the regime shift in the respective ecosystems, the periodic input of nutrients and precipitation due to the seasonal cycle has been neglected, but could have an important impact too. The same argument applies to the analysis of physical processes in the climate system driven by periodic or quasi-periodic changes in the orbital parameters of the sun leading to a variation in solar insolation with periods of about $23,000$, $44,000$ or $100,000$ years, the well-known Milankovitch cycles \cite{Imbrie1992}. Particularly the latter are assumed to be the major drivers for the development of the ice-ages before the Holocene \cite{Paillard2006}. Recently, these periodic forcings affecting the Albedo of the earth are studied to evaluate the impact of this variation on the Arctic and Antarctic ice cover \cite{Notz2009,Eisenman2009}.

In this paper we focus on tipping points and regime shifts of periodically forced systems. In this class of systems, abrupt changes of the dynamics at critical transitions do not occur between steady states (equilibria), but between oscillating states (limit cycles). Though at first sight, the hysteresis curves which are usually drawn to discuss critical transitions look similar, however, the dynamics is quite different. For this scenario we show that a transient structure, a so-called channel, occurring in the system's state space beyond the tipping point, creates a short-term dynamical regime with specific properties which attenuates the criticality of the transition. The smoothing of the transition is demonstrated by computing a finite-time measure of the twisting behavior (rotation property) of the state space surrounding trajectories while inside the channel domain. This measure indicates that, the trajectories passing the channel beyond the tipping point have residual system properties of the limit cycle destroyed at that tipping point. Hence, the channel acts as a ``ghost" of the destroyed limit cycle, retaining system trajectories in a very similar fashion. This fact is shown by statistical analysis of the intervals of time spent by noisy trajectories in the neighborhood of the limit cycle (pre-tipping) and the channel (post-tipping). Furthermore, we attribute to the ghost state the inconclusive diagnostic provided by variance and autocorrelation in anticipating tipping points.

\begin{figure}[!htp] \centering
\includegraphics[width=8.5cm,height=6.5cm]{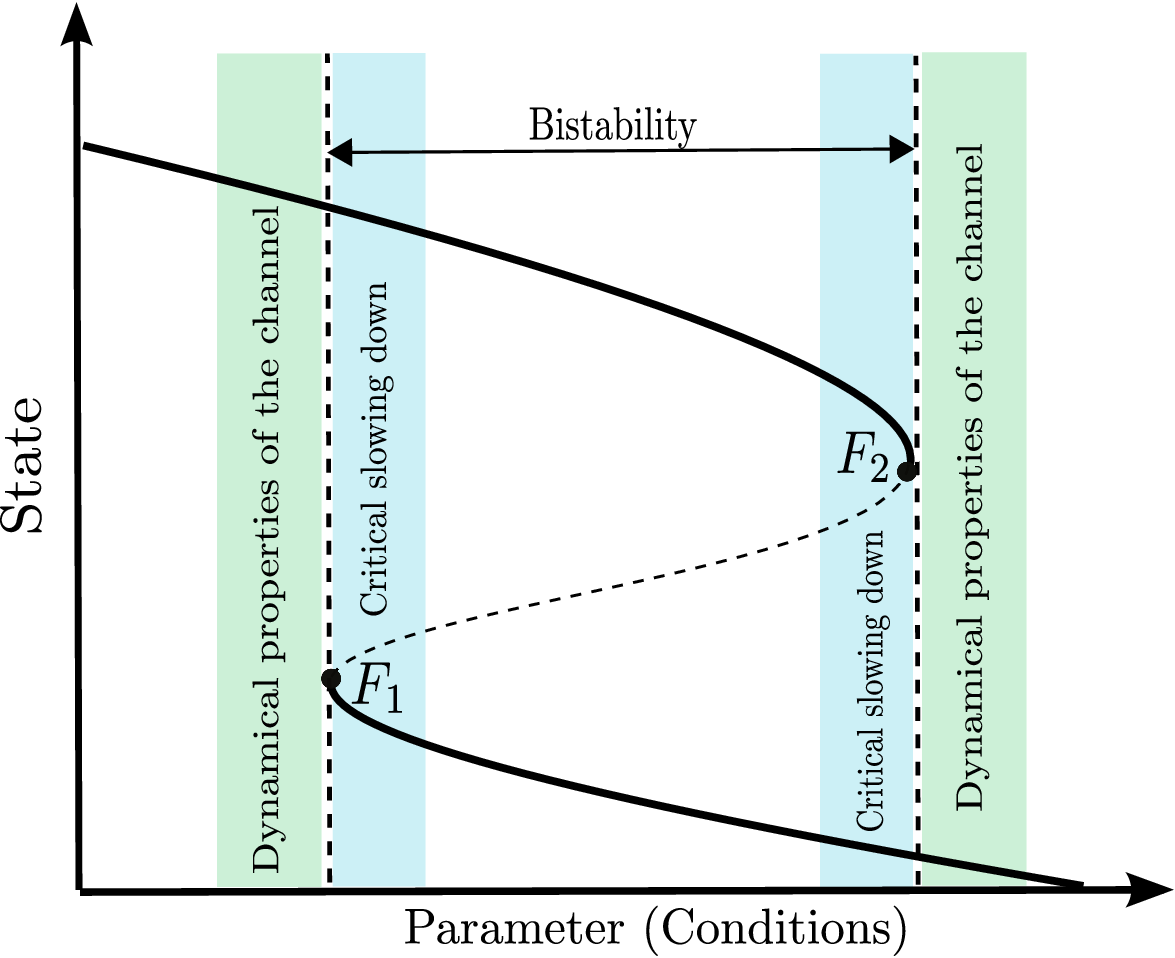}
\caption{Schematic of a bifurcation diagram illustrating the coexistence of two stable states (black curves) with an unstable saddle (dashed curve). The coexistence defines a bistable parameter region bounded by two parameters values corresponding to fold bifurcations, $F_1$ and $F_2$, which are the tipping points. The parameter regions where the critical slowing down phenomenon happens (which provides early-warnings to predict tipping points) are indicated in blue color. The green color regions indicate parameters where a channel associated to Fold bifurcations is formed.}
\label{schematic1}
\end{figure}

Let us now indicate the differences in the dynamical approach to deal with limit cycles instead of asymptotic equilibria. Figure~\ref{schematic1} shows the typical bifurcation diagram used to explain the appearance and disappearance of the coexistence of two alternative states. In contrast to the usual diagrams, the lines denote here one point of a limit cycle instead of stationary points. Therefore, the y-axis does not show one coordinate of the stationary state of the system, but one coordinate of the Poincar\'e section, a special construction which is very useful in analyzing periodic solutions of nonlinear dynamical systems (see Methods section). The Poincar\'e section in a periodically forced system defines a stroboscopic map in which the system is always analyzed at the same phase of the forcing, i.e. at times $t, t+T, t+2T,...$. Hence, a limit cycle corresponds to a fixed point in this stroboscopic map which makes the similarity between Fig. \ref{schematic1} and the well-known sketches of bistability in the stationary case obvious. 

There are two saddle-node or fold bifurcations of limit cycles denoted by $F_1$ and $F_2$, at which two limit cycles, a stable and an unstable one, emerge or disappear. Crossing those tipping points the dynamics will change dramatically. Increasing the parameter value $p$, the continuation of the limit cycle on the upper branch will stop abruptly at $F_2$ and switch to a periodic behavior corresponding to the lower branch, while decreasing the parameter $p$ and continuing the lower branch will result in a transition to the upper one at the critical threshold $F_1$. Our main focus lies in the analysis of the regions around those critical transitions. Firstly, we address the question to what extent the usual criteria of critical slowing down and flickering will signal the approach to the transition (blue region). Secondly, we show that the critical transitions are hidden due to particular structures in state space, so-called channels, which appear in the neighborhood of the fold bifurcations of limit cycles, preventing a clear identification of the critical transitions.

\section{Results}

To illustrate our results, we employ a paradigmatic model system, the well-known Duffing oscillator \cite{Holmes1979} and apply a periodic forcing with amplitude $A$ and frequency $\omega$. In mathematical terms, this simple dynamical system  reads: 

\begin{eqnarray}
    \ddot{x}+d \dot{x}-x+x^3 = A \cos(\omega t) +\sigma \xi(t) 
  \label{ecomodel}
\end{eqnarray}
The parameter $d$ is the amplitude of the system damping. The parameter $\sigma$ controls the noise intensity given by the stochastic forcing $\xi(t)$.  The function $\xi(t)$ represents the usual zero mean and unit variance noise with $\langle \xi(t)\xi(t')\rangle=\delta(t-t')$. We use a fourth-order Runge-Kutta method to integrate Eq.~(\ref{ecomodel}), in this process, the time is measured as a function of the period of the external forcing, i.e, $T=2\pi/\omega$. 

%\new{The time evolution of the solutions of Eq.~\ref{ecomodel} is here measured in units of the period of the system forcing which is given by $T=2\pi/\omega$.} 

\begin{figure}[!htp] \centering
\includegraphics[width=10cm,height=8cm]{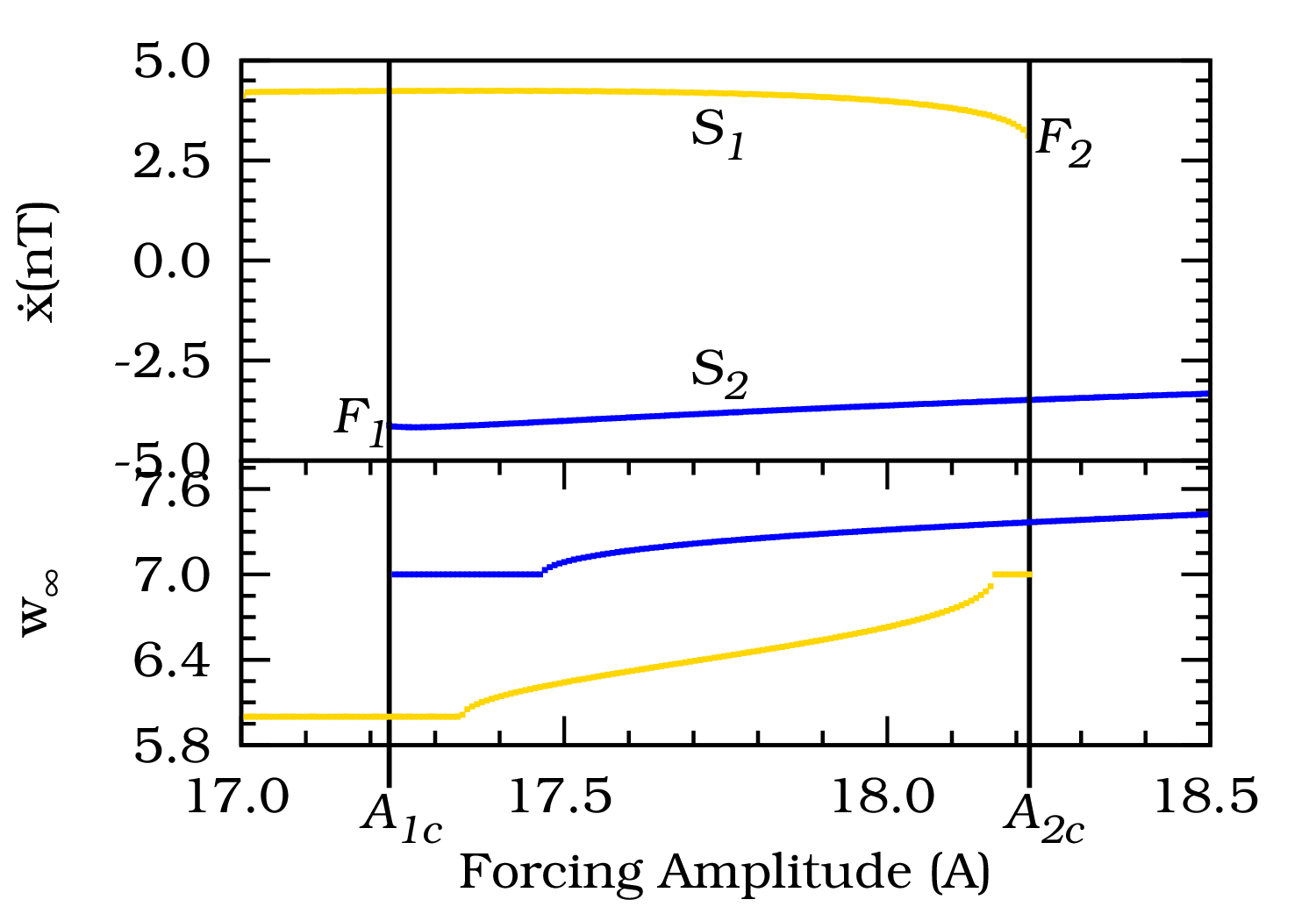}
\caption{(Upper) Bifurcation diagram of the noise-free ($\sigma=0$) Duffing oscillator showing a bistability of limit cycles. The different colors, blue and yellow, represent each limit cycle, $S_2$ and $S_1$, respectively. The state variable $\dot{x}(nT)$ is the $T$-shift map of the limit cycle variable, $\dot{x}$. The points $F_1$ and $F_2$ mark the parameters where catastrophic shift occurs, $A_{1c}=17.2295$ and $A_{2c}=8.2250$ are the corresponding critical parameter values. The other system parameters are settled in $d=0.3$, $\omega=0.5$. (Bottom) The asymptotic generalized winding numbers, $w_{\infty}$ , of each limit cycle in the parameter interval. The colors correspond to the respective limit cycles.}
\label{figure1}
\end{figure}

In a certain parameter range, the system described by Eq. (\ref{ecomodel}) exhibits a generic scenario of bistability between two different limit cycles, i.e. two stable periodic solutions exist separated by an unstable one of saddle character. The corresponding bifurcation diagram is shown in the upper panel of Fig. \ref{figure1}. Though this diagram looks very similar to the general diagram depicted in Fig. \ref{schematic1}, it shows only a Poincaré section of the stable limit cycles occurring for the system described by Eq. (\ref{ecomodel}). To characterize those limit cycles in more detail, we compute the generalized winding numbers (GWN) for each of them along the bistable parameter range, the results are depicted in the bottom panel of Fig. \ref{figure1}. In this panel, the GWN is represented by $w_{\infty}$, this measure quantifies the asymptotic twisting of the local in neighborhood of limit cycles, a better description of this measure is given in the {\bf Methods}.

\begin{figure}[!htp] \centering
\includegraphics[width=10cm,height=9cm]{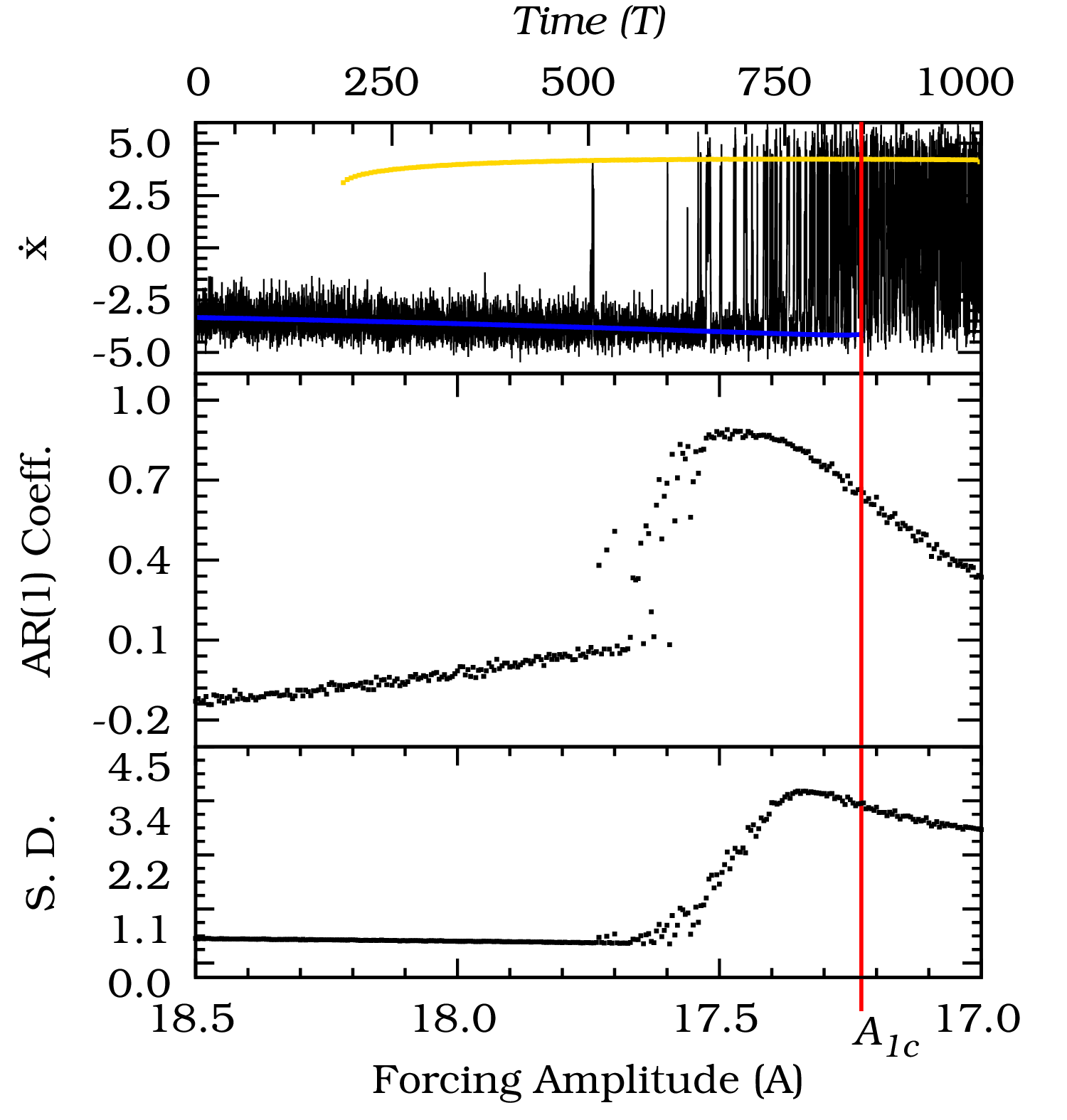}
\caption{(Upper) The black line indicates the time evolution of a noisy trajectory for the noise level fixed at $\sigma = 0.02$. The forcing amplitude, $A$, is varied linearly through the bifurcation diagram which the $T$-shift map of the noiseless asymptotic limit cycle is represented by the colors blue and yellow (Same bifurcation diagram of Fig.~\ref{figure1}(Top)). (Middle) The black points represent the autocorrelation coefficient at lag-$1$ as the forcing amplitude approaches the critical value $A_{1c}$. (Bottom) The black points represent the standard deviation of the average value of the noisy time series.}
\label{figure2}
\end{figure}

The bifurcation diagram of Fig.~\ref{figure1}(Upper) shows the dependence of the noise-free Duffing oscillator on the forcing amplitude $A$. Two stable limit cycles, $S_1$ (blue) and $S_2$ (yellow), coexist for a range of parameters bounded by two fold bifurcations of limit cycles at the points $F_{1}$ and $F_{2}$ (tipping points). So, the system is subject to catastrophic shifts, tipping points, as the parameter $A$ reaches the points $A_{1c}$ or $A_{2c}$. Let us now check whether the autocorrelation coefficient at lag-$1$ and the variance of the system indicate the approach of the critical transition and can serve as early warnings. To this end, we apply now noise to the system and show the resulting behavior in Fig. \ref{figure2}. We focus our analysis on the parameter region close to $A \sim A_{1c}$. Hence, in Fig.~\ref{figure2}, we reverse the $x$-axis to better investigate the critical transition at the parameter $A_{1c}$. In Fig.~\ref{figure2} (Upper panel), the black line indicates the time evolution of a noisy trajectory with the parameter $A$ varying in the same interval of the bifurcation diagram also indicated in this panel. In Fig.~\ref{figure2} (Middle panel), we show the autocorrelation at lag-$1$, a measure that usually increases with the approach of critical transitions of systems in equilibrium. In this panel, for limit cycles, we verify a sudden increase in the autocorrelation coefficient as soon as the noisy trajectory starts flickering between the stable limit cycles. But subsequently, it decreases as the system approaches the critical threshold and does not exhibit any significant change as the critical threshold $A_{1c}$ is passed. Similar behavior is observed for the standard deviation of the noisy trajectory, shown in Fig.~\ref{figure2} (Bottom panel). Therefore, we find that the usual indicators of critical transitions between equilibria may not work for limit cycles. Instead we observe a continuous decreasing of the autocorrelation coefficient and the variance, not suitable to serve as an early warning signal. To explain this behavior, we investigate in more detail the state space structure occurring for parameters succeeding the fold bifurcation at $A_{1c}$.

\begin{figure}[!htp] \centering
\includegraphics[width=16cm,height=6cm]{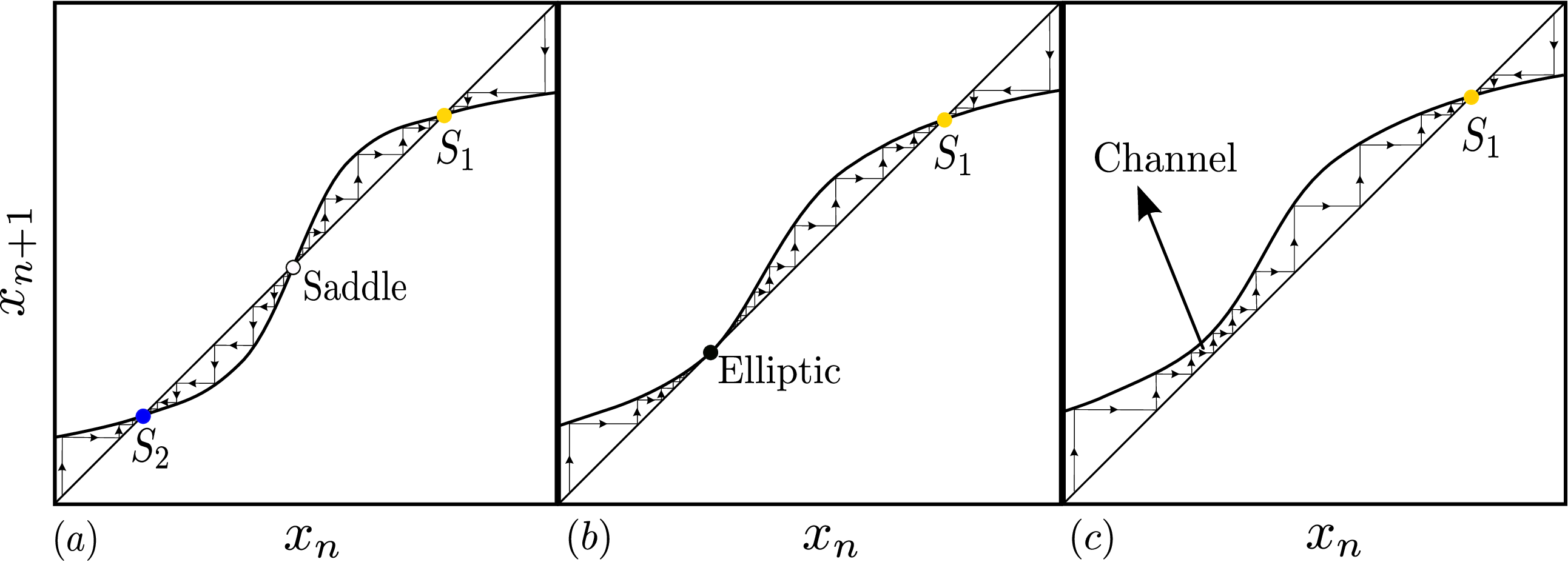}
\caption{($a$) Two fixed points of the node type ($S_1$ and $S_2$) are coexisting with a saddle. The black arrows indicate how initial conditions dynamically behave in the neighborhood of each fixed points. ($b$) As a system parameter is varied the node $S_2$ and the saddle collide forming an elliptic point $E$ (saddle-node or fold bifurcation). ($c$) As the parameter crosses the critical bifurcation parameter, the initial conditions (arrows) that used to belong to the attraction domain of the extinct node $S_2$ are now converging to the node $S_1$ through a channel formed in the mapping.}
\label{schematic2}
\end{figure}

Fold bifurcations of limit cycles are accompanied by the formation of channels in state space through which the trajectory has to go after entering it. To illustrate this behavior which has been first described by Manneville \cite{Manneville1979, Pomeau1980} in the context of intermittency in turbulence, we show in Fig. \ref{schematic2} the generic principle behind the formation of that channel. As outlined above, a limit cycle corresponds to a fixed point in the Poincar\'e section. In our case, one point $x_n$ in the Poincar\'e section is mapped onto the next point $x_{n+1}$ by mapping the limit cycle stroboscopically every period of the forcing, so the fixed points are mapped onto the diagonal $x_{n+1}=x_n$ of the diagram $x_{n+1} \times x_n$ shown in Fig. \ref{schematic2}. In the bistable region we have three fixed points, two stable and one unstable separating the former two (Fig. \ref{schematic2}, left panel). When the fold bifurcation is reached the stable and the unstable limit cycle merge and form an elliptic point (middle panel), while beyond the fold bifurcation a channel appears in state space through which the trajectory moves when it comes close to the region in state space where previously the two limit cycles have been located. Without noise, the trajectory would finally converge to the only stable limit cycle left in the system, denoted by $S_1$. Due to the noise, the trajectory is kicked back to the channel from time to time and moving through it again and again. As a consequence of this behavior we observe even beyond the fold bifurcation, that the dynamics returns to the ``ghost" of the limit cycle manifested as the channel. The resulting dynamics contains phases where the trajectory is close to the ``ghost" and phases where is comes close to the only stable limit cycle but kicked away again by the noise. This way, the flickering dynamics goes on even though the system is well beyond the critical transition. For the very same reason the widely used indicators for critical transitions such as lag-$1$ autocorrelation function, variance as well as flickering can not signal the approach to the critical transition and the shift or tipping points happens with no warning. In our case, the characteristics of the critical slowing down indicators resemble the case of a smooth transition as analyzed in \cite{Kefi2013}. Additionally, we note that the dynamics before and beyond the critical transition is essentially the same, characterized by the hopping between the two stable states before and between the single stable state and the ``ghost" beyond the tipping point. This behavior is generic and will occur for all fold bifurcations of limit cycles forming a channel after the bifurcation.

\begin{figure}[!htp] \centering
\includegraphics[width=8.5cm,height=6.5cm]{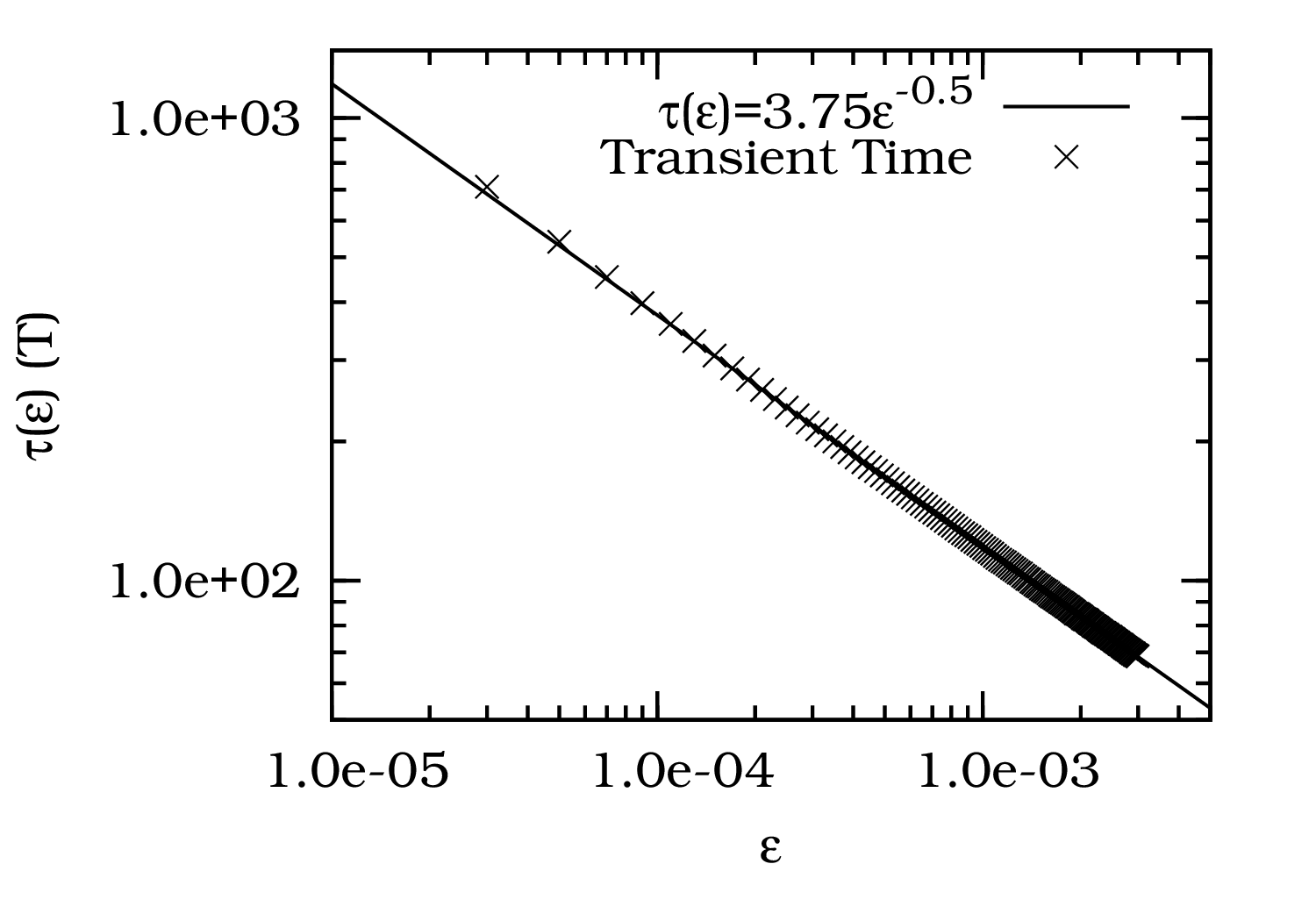}
\caption{The average transient time $\tau(\varepsilon)$, measured in units of the systems forcing $T$, for initial conditions to leave the channel and arrive at the remaining attractor as a function of the distance $\varepsilon$ from the critical parameter point $A_{1c}$. Crosses represent the numerical results and the black straight line a power-law fitting.} 
\label{figure3}
\end{figure}

Let us now discuss the post-tipping behavior in more detail. To demonstrate that indeed the channel is the most essential structure in the state space deforming the noisy system beyond crossing the critical threshold, we analyze the scaling behavior of the length of the transient time with the distance from the threshold, Therefore, we define $\varepsilon$ as a parameter measuring the distance from the critical threshold $A_{1c}$, i.e., $\varepsilon = A_{1c}-A$. Then, as a function of this distance $\varepsilon$, we measure the transient time, $\tau(\varepsilon$), for trajectories to reach the remaining stable limit cycle (yellow). For these trajectories, we choose a set of initial conditions in the state space region previously occupied by the basin of attraction of the limit cycle $S_2$ (blue) destroyed in $F_1$. In Fig.~\ref{figure3}, we show the results for an ensemble of $300$ random initial conditions for each distance $\varepsilon$. We find the time that trajectories spend to leave the channel scales as a power-law with the distance $\varepsilon$. The characteristic exponent is equal $0.5$, and hence, it corresponds to the value known from the type $I$ intermittency \cite{Manneville1979,Pomeau1980}. Hence, the characteristic time $\tau(\varepsilon)$ enables us to quantify the influence of the channel in the time evolution of trajectories as a function of the parameter distance $\varepsilon$.

\begin{figure}[!htp] \centering
\includegraphics[width=9cm,height=7cm]{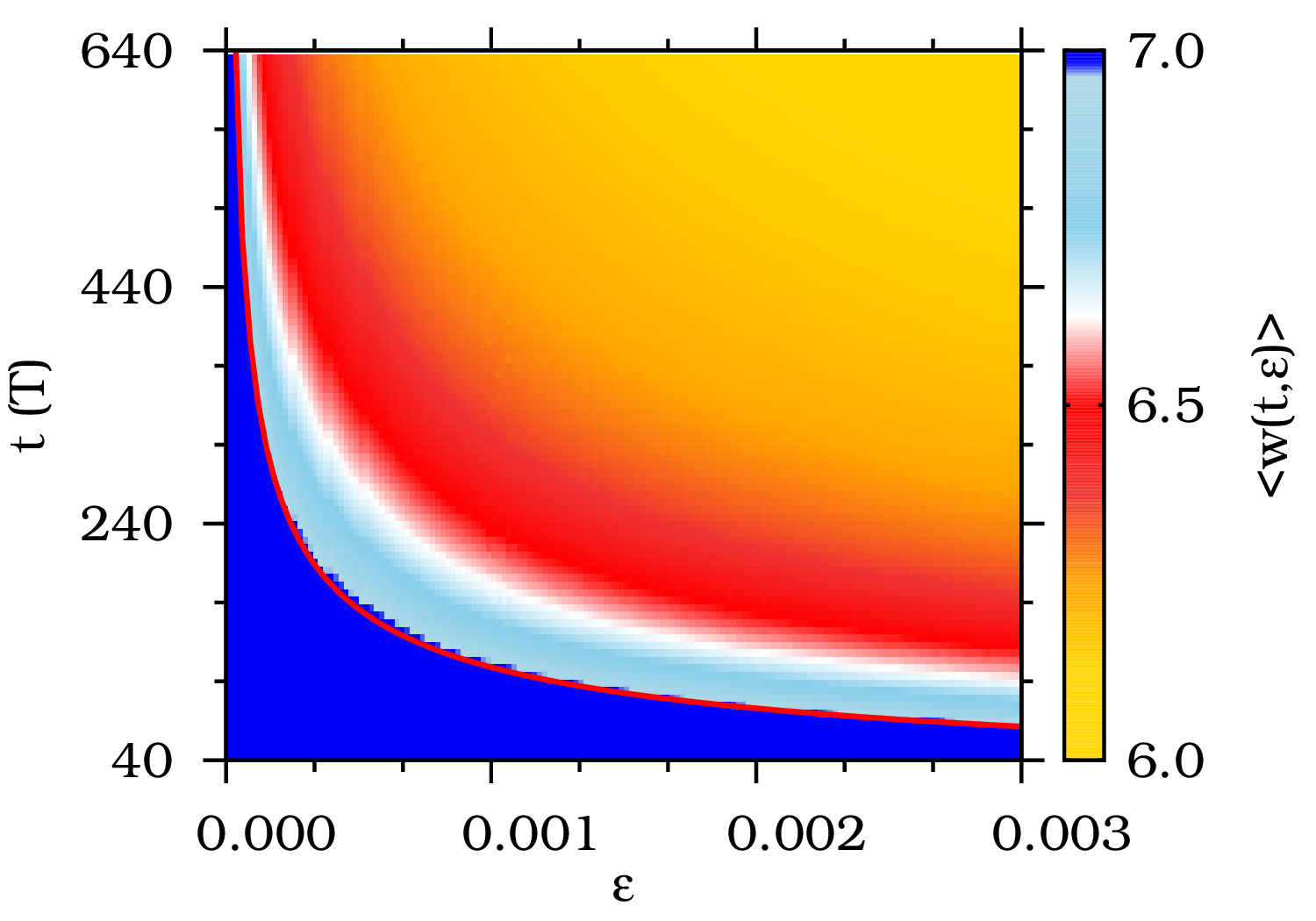}
\caption{A diagram, $t \times \varepsilon$, i.e., the time evolution, $t$, for the parameter distance $\varepsilon$. The color scale represent the finite-time winding numbers $\left\langle w(t,\varepsilon) \right\rangle$. The red line indicates the power-law function, $\tau(\varepsilon)=3.75\varepsilon^{-0.5}$, adjusted in Fig.~\ref{figure3} for the time spent by trajectories to cross the channel.}
\label{figure4}
\end{figure}

This equality verifies that trajectories starting in the former basin of attraction of the limit cycle destroyed in the fold bifurcation are in fact being trapped in the channel associated to this bifurcation for a characteristic time, $\tau(\epsilon)$. In order to obtain the twisting behavior of trajectories just during the time trapped in the channel, we introduce a finite-time version of the winding number (FTWN) represented by $\left\langle w(t,\varepsilon) \right\rangle$. A complete description of this definition is given in section {\bf Methods}. In the diagram shown in Fig.~\ref{figure4}, the color code indicates the FTWN given by $\left\langle w(t,\varepsilon) \right\rangle$, in the $y$-axis we represent the time evolution, $t$, in units of the period of the forcing, while in the $x$-axis we show the distance $\varepsilon$ from the bifurcation point. The red line represents the function $\tau(\varepsilon)$ obtained from the adjustment in Fig.~\ref{figure3} for the characteristic time for trajectories to cross the channel. We observe that regardless of the parameter distance $\varepsilon$, the finite-time winding number has a defined value equal to $7.0$ (blue in Fig.~\ref{figure4}) for times lower than the corresponding $\tau(\varepsilon)$. Hence, from Fig.~\ref{figure4}, we conclude that the post-tipping trajectories, while crossing the channel, conserve the twisting behavior (rotation properties) of the stable limit cycle destroyed in the tipping point.

\begin{figure*}[!htp] \centering
\includegraphics[width=17cm,height=6.5cm]{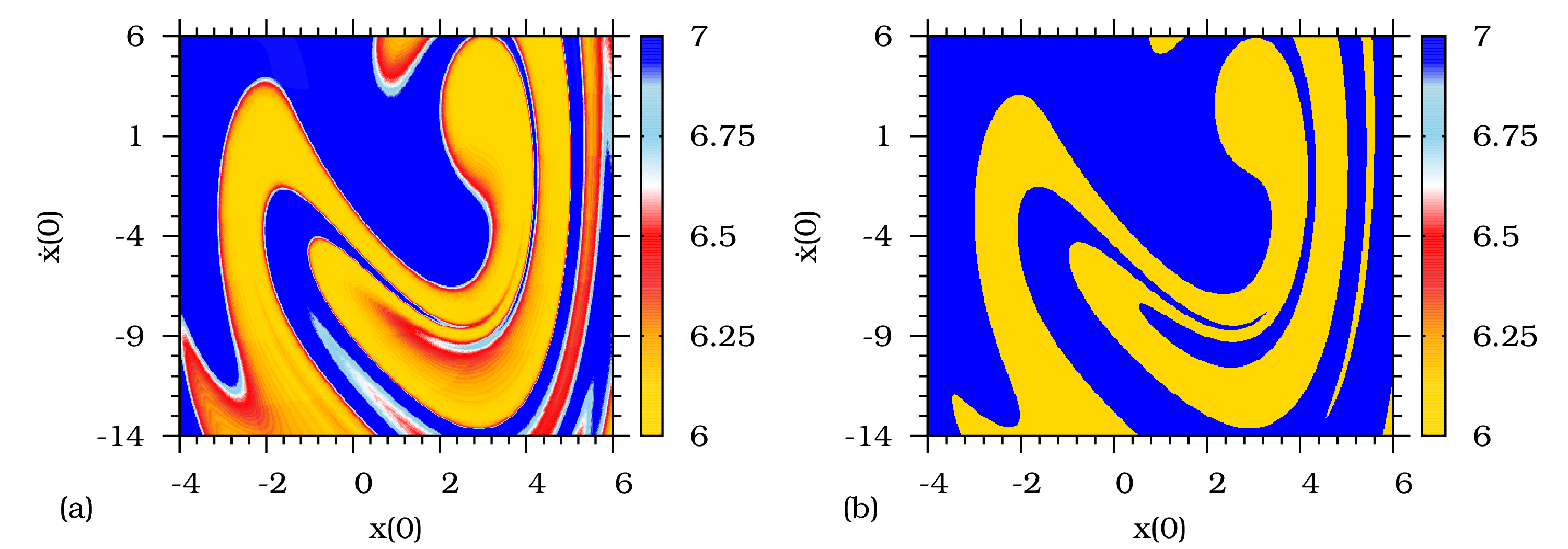}
\caption{(a) Sets of initial conditions leading to the dynamical channel (blue) and to a stable state (yellow), the color scale indicate the FTWN of trajectories corresponding to each initial condition. The system parameters are fixed in $d=0.3$, $\omega=0.5$, and $A=17.22$. The time used to compute the FTWN is fixed at $\tau(0.0095)=33.34$ ($T$). (b) Generalized asymptotic winding number (GWN) for The parameter $A=17.30$ (bistable region). The GWN is computed for the same sets of initial conditions of ($a$).}
\label{figure5}
\end{figure*}

In the following, we confirm the existence of the residual twisting behavior of the destroyed limit cycle by obtaining the FTWN of sets of initial conditions crossing the channel. Firstly, we choose the parameter $A$ such that the dynamics will take place in the channel, i.e., $A_{1c}$ minus a small distance $\epsilon=0.0095$, then we compute the FTWN during the time $\tau(0.0095)=33.34$ ($T$) for a grid of initial conditions. Attributing different colors to the FTWN obtained for the trajectories corresponding to each initial condition, we clearly distinguish, in the grid of Fig.~\ref{figure5}(a), two types of dynamic behavior. (i) the FTWN corresponding to trajectories that cross the channel (initial conditions in blue, fast twisting) and (ii) the the trajectories converging directly to the remaining stable state (initial conditions in yellow, slow twisting). In order to compare the twisting properties of the channel, measured by FTWN, to the twisting behavior around the stable states in the bistable region, we characterize the twisting around the two stable states, $S_1$ and $S_2$, by the asymptotic generalized winding number (GWN). Fig.~\ref{figure5}(b) shows those twisting properties of trajectories starting on the same grid as in Fig.~\ref{figure5}(a) but computed by the asymptotic (GWN) for a forcing amplitude $A$ before the tipping. We notice the similarity between twisting of trajectories crossing the channel beyond tipping (blue in Fig.~\ref{figure5}(a)) and around the stable state before tipping (blue in Fig.~\ref{figure5}(b)).

\begin{figure*}[!htp] \centering
\includegraphics[width=17cm,height=6.5cm]{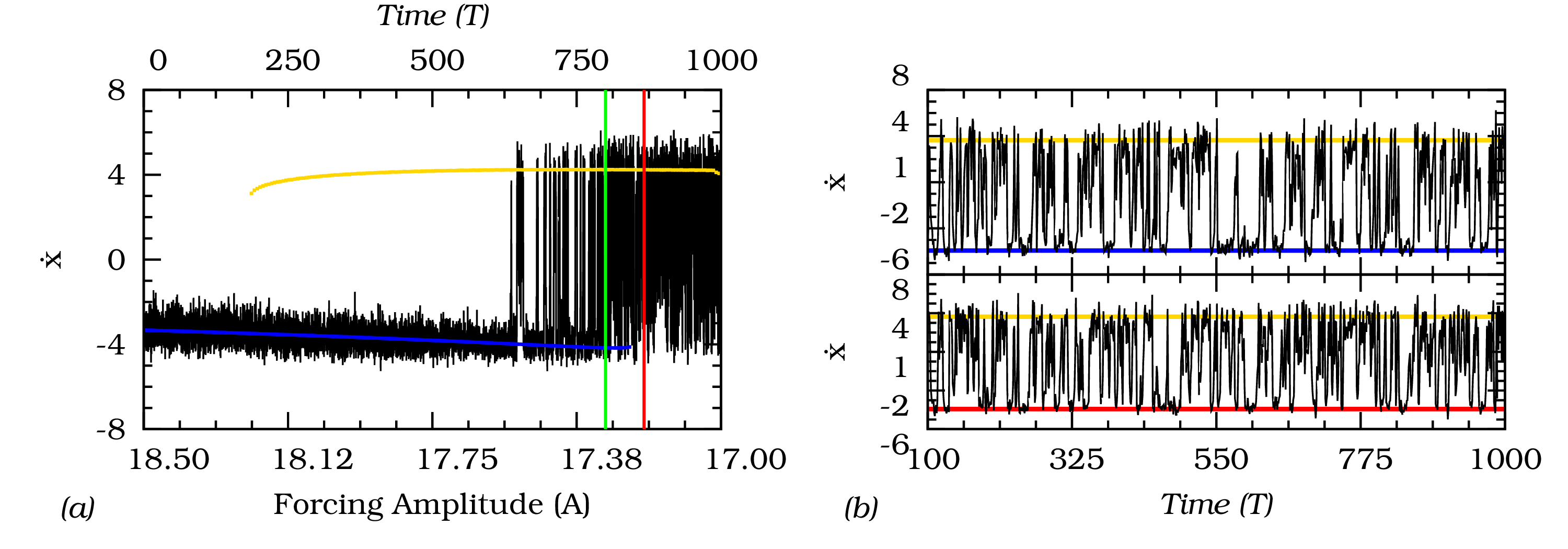}
\caption{($a$) The black line indicates the time evolution of the noisy trajectory with the parameter $A$ varying in accordance to the bifurcation diagram as shown. The other parameters are set in $d=0.3$, $\omega=0.5$, and the noise intensity fixed at $\sigma=0.025$. ($b$) In the top is the temporal evolution of the perturbed trajectory for parameter $A$ fixed at $17.30$ (green line in Figure~\ref{figure6}($a$). The yellow and blue lines represent the position of each stable limit cycle in the pre-tipping region. In the bottom is the temporal evolution of the noisy trajectory for the parameter $A$ fixed at $17.22$, (red line in  Figure~\ref{figure6}($a$)). The red line in the bottom panel represents the position of the state space channel visited by the noisy trajectory.}
\label{figure6}
\end{figure*}

To illustrate further that observations of the system`s trajectories are insufficient to determine whether the system is bistable (pre-tipping) or has a dynamical channel (post-tipping), we show in Fig.~\ref{figure6}($a$) the temporal evolution of a trajectory of the noisy the Duffing oscillator as the parameter $A$ increases with time in the same interval as in Fig.~\ref{figure6}($a$). We notice that, even after the limit cycle $S_2$ marked in blue disappears in $F_1$, the noisy trajectory (black line) is still flickering into the state space region previously occupied by the extinct limit cycle around $\dot{x}=-4$. This becomes even more obvious when comparing two noisy trajectories with fixed forcing at an amplitude $A$ in the bistable region (pre-tipping) to a trajectory with a forcing amplitude beyond the tipping point (red line in Fig.~\ref{figure6}($a$). In a statistical sense those two trajectories are indistinguishable, indicating that the pre-tipping and the post-tipping behavior are very similar, with flickering between two distinct state space regions of $S_1$ and $S_2$ or the ``ghost" of $S_2$ respectively.

\begin{figure*}[!htp] \centering
\includegraphics[width=17cm,height=10cm]{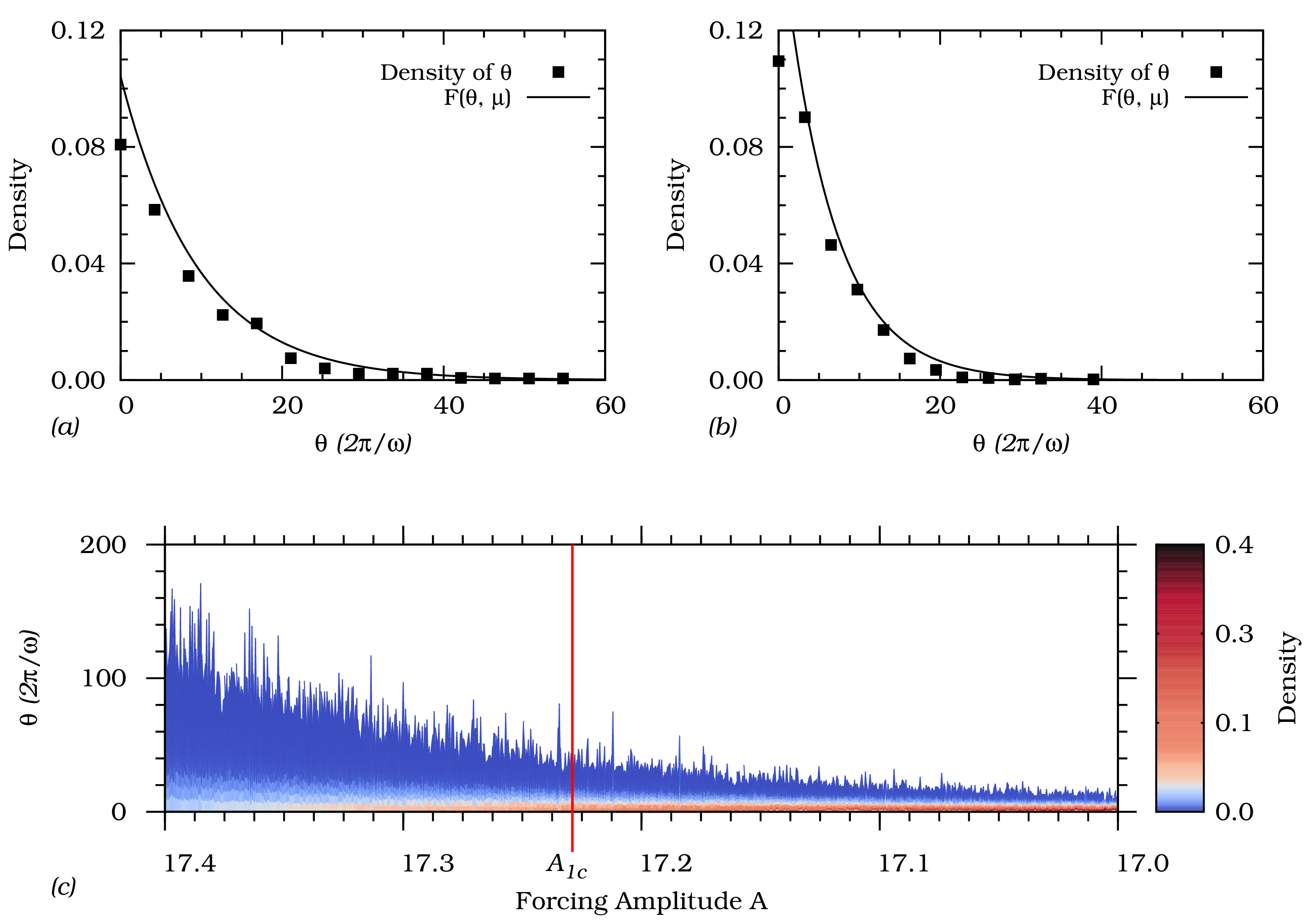}
\caption{($a$) The black rectangles indicate the distribution of normalized intervals of time, $\theta$, that a trajectory elapses around a stable state. The full line indicates the exponential distribution for which the normalized PDF is represented by $F(\theta$, $\mu)$. The model parameters are set to $d=0.3$, $\omega=0.5$, and $A=17.30$. The noise level is $\sigma=0.025$. The mean value of the PDF is $\mu=9.60$. ($b$) The system parameters of the previous, except for the parameter $A$ fixed at $A=17.22$ corresponding to the channel region. For this case, the mean value of the data is $\mu=6.25$. ($c$) The intervals of time, $\theta$, as function of the system parameter $A$. The color code indicates its probability density. The red line marks the parameter $A_{1c}$ corresponding to the tipping point.}
\label{figure7}
\end{figure*}

As a consequence, time series as the main window to observations in nature, would show the flickering phenomenon before and beyond the tipping points making the transition in the observed data to appear smooth instead of abrupt. In order to verify this statement, we investigate the intervals of time, $\theta$, that a noisy trajectory elapses in the neighborhood of the stable state (before the tipping point), and in the channel (beyond the tipping point). The idea behind this study is to extend the notion of escape times to characterize the dynamics beyond the tipping point. In bistable systems one usually computes the mean escape time or mean first passage time to identify the stability of each stable state in a stochastic sense. While for systems possessing a double well potential, it is possible to compute those escape times analytically \cite{Hanggi1990}, one has to rely on numerical estimations for arbitrary multistable systems \cite{Kraut1999}. Though the vast majority of nonlinear dynamical systems do not possess a potential, the scaling of the escape rates remains valid. Specifically, in Figure \ref{figure7}, we obtain the distribution of time intervals spent by trajectories in the neighborhood of the stable states and in the channel. The time interval, $\theta$ is also expressed in units of the period of the forcing $T$.

Specifically, we show in Fig.~\ref{figure7} the distributions of the time intervals spent by trajectories in the neighborhood of the stable limit cycle $S_2$ which will go extinct at the tipping point (Fig.~\ref{figure7}($a$)) with the distribution of those time intervals spent in the neighborhood of the ``ghost" of $S_2$ beyond the tipping point (Fig.~\ref{figure7}($b$)). Both distributions are exponential distributions, so that the probability density function can be approximated by 
\begin{equation}
F(\theta, \mu) = \frac{1}{\mu}\exp{\left(-\frac{\theta}{\mu} \right)}
\label{GEV}
\end{equation}
where $\mu=\left\langle \theta \right\rangle > 0$ is the mean value of the distribution. While in the bistable parameter region the mean time spent close to the limit cycle $S_2$ is $\left\langle \theta \right\rangle  \approx 10$ periods of the forcing, it is only slightly shorter ($\left\langle \theta \right\rangle  \approx 6$ periods of the forcing) beyond the tipping point. However, the density function for the dynamics close to the channel is narrower and higher than in the bistable region, indicating that the shorter intervals of time are more frequent. Hence, even for the parameter $A$ lower than $A_{1c}$ (beyond tipping point), the frequency with which trajectories visit the neighborhood of the extinct state is not zero, i.e., the flickering phenomenon still occurs after the tipping point. It means that even after the extinction of the limit cycle in $A_{1c}$, the state space channel keeps retaining trajectories, avoiding their abrupt definitive transition to the unique survival stable state.

To emphasize that, the characteristics of the dynamics changes smoothly and not abrupt when crossing the tipping point, we show in Fig.~\ref{figure7}($c$) the changes in the distribution function when decreasing the forcing $A$ from the bistable to the monostable region. In this figure, the color code indicates the probability densities for each parameter $A$ shown in the $x$-axis. We observe that the distribution of time intervals smoothly changes as the tipping point is approached and passed, indicating that there is no abrupt transition crossing the threshold. However, as the parameter $A$ is passed through the tipping point, a considerable increasing in the density of time interval values around the expected value is observed making the distribution narrower for parameters well beyond the tipping point.

\section{Discussion}

In summary, we address tipping points of systems subjected to periodic external forcing. The asymptotic solutions of this class of systems inherently settle into oscillating stable states (limit cycles), a more complex dynamics than the stable steady states (equilibria) for which the tipping points have been widely studied. In nature, the most noticeable occurrences of such oscillating attractors are found in ecology and climate sciences where periodic and quasi-periodic variabilities arise from external factors such as seasonality and astronomical forcing. Here, for a generic periodically forced system that generates such oscillations, we consider the typical hysteretic scenario to investigate tipping points, i.e., a bistable parameter region where the tipping is represented by fold bifurcations of limit cycles rather than steady states. As the parameters are varied and the system reaches a fold bifurcation, in which a stable limit cycle is destroyed leaving a transient structure, a so-called channel, in the state space of the system. Hence, for parameters beyond this tipping point, the channel gives rise to a short-term dynamics which possesses similar properties than the destroyed limit cycle and can therefore be attributed to a "ghost" of the latter. We find that a residual dynamical property of the limit cycle destroyed in the tipping point, namely its twisting behavior, occurs in the short-term dynamics for parameters in the post-tipping region. This finding indicates that the short-term behavior carries dynamical information of the destroyed oscillating stable state. 

For system parameters fixed in the post-tipping region, we obtain the time evolution of the system subject to a stochastic noise. With this, we show that the ``ghost" attractor retains systems trajectories in a very similar fashion of the stable limit cycle destroyed in the tipping point. Additionally, by obtaining the statistics of the time intervals that noise trajectories spend in the neighborhood of the stable limit cycle and in the neighborhood of the ``ghost", we find that the PDFs of waiting times in both regions have the same exponential profile and do not differ much in their expected values. Therefore, the ``ghost" dynamics plays an essential role in attenuating the critical transition in a way that it may be seen as a smooth transition when trying to diagnose  it from real-world data. Hence, none of the well-known methods like autocorrelation function, variance or flickering are suitable to identify this particular transition properly. The emergence of the "ghost" masks the transition until the system is well beyond the tipping point and makes it to appear smooth instead of catastrophic.

\section{Methods}
\label{methods}
\subsection*{Poincar\'e Section}

As we consider systems whose asymptotic behavior are {\it limit cycles}, the final dynamics are oscillations rather than equilibria. Bifurcation analysis are performed by defining a Poincar\'e section, which usually is a hyper-surface arranged transversally to the limit cycle, where the whole system dynamics is described by a discrete system. Letting $f$ be the function that describes the intersections of the limit cycle with the section, for trajectories in three dimensional space, results that $(x_{1(n+1)},x_{2(n+1)})=f(x_{1n},x_{2n})$, where $x_{1n},x_{2n}$ are the coordinates of the $n^{th}$ crossing. Consequently, on the surface of section, limit cycles are represented by {\it fixed points} of $f$. Then, states, as shown in bifurcation diagrams such as of Fig.~\ref{schematic1}, are defined in the surface of section, and in case the section is chosen to be a plane, they are denoted by the ($x_{1}$,$x_{2}$) plane coordinates. 

For the Duffing oscillator described by Eq.~(\ref{ecomodel}), a suitable Poincar\'e section is the so-called $T$-shift. The dynamics over the section is represented by discrete variables ($\dot{x}(nT)$,$x(nT)$) defined as the solution pair ($\dot{x}$, $x$) collected every period, $T=2\pi/\omega$.

\subsection{Fold Bifurcation of Limit Cycles and State Space Channel}

\begin{figure*}[!htp] \centering
\includegraphics[width=17cm,height=5cm]{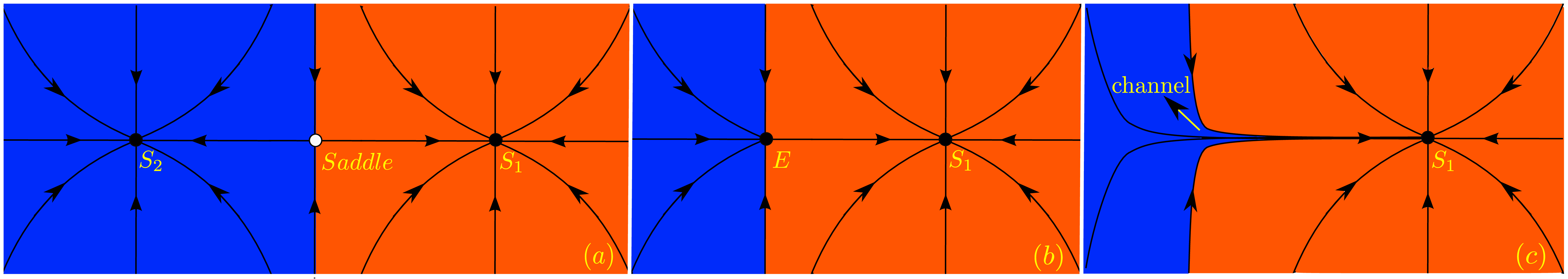}
\caption{Two dimensional representation of the bifurcation scenario. ($a$) Two fixed points of the node type ($S_1$ and $S_2$) are coexisting with a saddle. The black lines indicate how initial conditions dynamically behave in the neighborhood of the fixed points. The colors indicate the domain of attraction of each node. ($b$) As a system parameter is varied the node $S_2$ and the saddle collide forming an elliptic point $E$ (saddle-node or fold bifurcation). ($c$) As the parameter crosses the critical bifurcation parameter, the initial conditions that used to belong to the attraction domain of the extinct node $S_2$ are now converging to the node $S_1$ through a channel formed in the plane.}
\label{schematic3}
\end{figure*}

In case of limit cycles, a risky bistable configuration occurs when two {\it stable limit cycles} are coexisting with one {\it unstable cycle of saddle type}. The emergence of a dynamical channel at this scenario can be described on a suitable Poincar\'e section. We show in Fig.~\ref{schematic3}($a$) that the stable limit cycles yield two fixed points of the {\it node type}, $S_1$ and $S_2$ in the surface of section, while the unstable limit cycle generates a fixed point of {\it saddle type}. The stable manifold of the saddle separates the initial conditions attracted by each node (blue and red in Fig.~\ref{schematic3}($a$)). As the control parameters are varied approaching the fold bifurcation that delimitate the bistability region, such as $F_1$ and $F_2$ in Fig.~\ref{schematic1}, one of the node fixed points approaches the saddle. When the system is set to the parameters at the fold bifurcation point, for instance, the node $S_2$ and the saddle collide, and they both disappear forming an elliptic fixed point denoted by $E$ in Fig.~\ref{schematic3}($b$). For post-bifurcation parameters, the schematic of the Poincar\'e section is shown in Fig.~\ref{schematic3}($c$), the node and the saddle no longer exist in the Poincar\'e section. However, trajectories starting in the space state region, which used to be the basin of attraction of the destroyed node $S_2$, converge to the remaining node but not before being attracted by the stable manifold of the unstable elliptic point. Effectively, the system's trajectory behaves as if there existed a channel constraining the trajectory and leading it to the remaining stable state.

The occurrence of these dynamical channels related to fold bifurcations of limit cycles has been first discussed by Pomeau and Manneville, and has been argued to be the mechanism responsible for the laminar phase in the type-$I$ intermittency scenario \cite{Manneville1979,Manneville1980,Pomeau1980}. However, in type-$I$ intermittency, chaotic bursts re-inject the trajectory in the dynamical channel. The trajectory spends long time intervals to cross the channel (the laminar phase) eventually escaping to the chaotic phase space region, producing another chaotic burst. In this work, there is no chaotic process to re-inject the trajectory into the channel, so we introduce a Gaussian noise which resets the trajectory to a random configuration belonging to the basin of attraction of the stable state extinct in the fold bifurcation. This procedure successively ejects the trajectory off the neighborhood of the survival stable state, forcing it to successively cross the channel along its time evolution. Regardless of the mechanism used to re-inject the trajectory through the channel, the time spent by trajectories to cross depends on the distance $\varepsilon$ from the fold bifurcation as \cite{Manneville1979, Manneville1980, Pomeau1980}:

\begin{equation}
 \tau(\varepsilon)\sim \varepsilon^{-\frac{1}{2}}.
\end{equation}

\subsection{Finite-Time Winding Number (FTWN)}

In general, the state space in the neighborhood of a limit cycle is affected by its presence, commonly, the limit cycle induces a twisting in its neighboring space. This twisting can be quantified by computing the so called {\it Generalized Winding Number} (GWN). Given a limit cycle $\gamma$ of the duffing oscillator described by Eq.~(\ref{ecomodel}), the GWN of $\gamma$ can be obtained by computing the average frequency $\Omega_{\infty}$ of the twisting that a neighbor trajectory $\gamma'$ performs around $\gamma$ \cite{Parlitz1985a,Medeiros2013}. Defining $\alpha(t)$ as the angle between $\gamma$ and $\gamma'$ over the $T$-shift Poincar\' e section, the frequency is given by:
\begin{equation} 
  \Omega_{\infty} =
  \lim_{t \rightarrow  \infty} {\frac{1}{t} \left|\int_0^t \!
    \dot{\alpha}(t') \, \mathrm{d}t'\right|} = 
  \lim_{t \rightarrow \infty} {\frac{|\alpha (t) - \alpha(0)|}{t}}.
  \label{eq:inffreq}
\end{equation}
And, the GWN is \cite{Parlitz1985a}:
\begin{equation}
  w_{\infty} = \frac{\Omega_{\infty}}{\omega},
  \label{eq:infwind}
\end{equation}
where $\omega$ is the frequency of the forcing in Eq.~(\ref{ecomodel}).

Hence, Eq.~(\ref{eq:inffreq}) and (\ref{eq:infwind}) allow us to compute the GWN of the two coexisting stable states of the bistable region (between $A_{1c}$ and $A_{2c}$ in Fig.~\ref{figure1}). For each stable state a GWN can be calculated considering sets of initial conditions belonging to the basin of attraction of each state. Since these states are attractors, trajectories naturally go to the neighborhood of the stable limit cycles, providing a GWN based on the local properties of the limit cycles. 

Notice that Eq.~(\ref{eq:inffreq}) is defined for an infinitely long time, in a manner that the main contribution to the averaged twisting frequency $\Omega_{\infty}$ comes from rotations of the asymptotic stable state. Hence, to obtain twisting properties of transitory structures, Eq.~(\ref{eq:inffreq}) has to be reformulated in a finite time version. So, we define a finite-time twisting frequency as:

\begin{equation} 
  \Omega_{t} =
   {\frac{1}{t} \left|\int_0^t \!
    \dot{\alpha}(t') \, \mathrm{d}t'\right|} =
     {\frac{|\alpha (t) - \alpha(0)|}{t}}.
    \label{eq:2}
\end{equation}
For short-term trajectories we have to consider possible deviations in the finite-time winding number for different initial conditions. Hence, we take an average over an ensemble of initial conditions, resulting in the follow definition for the finite-time winding numbers:
\begin{equation}
  w(t) =  \left\langle\frac{\Omega_{t}}{\omega}\right\rangle,
  \label{eq:3}
\end{equation}
where the brackets denote the average over the ensemble of trajectories. As long the time to cross the channel, $\tau$, is a function of the parameter distance $\varepsilon$, we represent the finite-time winding number as $\left\langle w(t,\varepsilon) \right\rangle$, i.e., also a function of $\varepsilon$ instead of only $t$

%

%\bibliography{tippinglib}

\section{Acknowledgements}

We would like to thank the partial support of this work by the Brazilian agencies FAPESP (processes: 2011/19296-1, 2013/26598-0, and 2015/20407-3), CNPq and CAPES. MSB acknowledges EPSRC Ref. EP/I032606/1.

\end{document}